\documentclass[preprint,showpacs,preprintnumbers,amsmath,amssymb]{revtex4}
\usepackage{graphicx}
\usepackage{dcolumn}
\usepackage{bm}
\usepackage{epsfig}
\begin{document}
\preprint{}
   \title{High Resolution Spectroscopy of the $^{12}_{\Lambda}$B
   Hypernucleus Produced by the (e,e$^{\prime}$K$^{+}$) Reaction}
\author{
T. Miyoshi$^{a}$,
M. Sarsour$^{b}$,
L. Yuan$^{c}$,
X. Zhu$^{c}$,
A. Ahmidouch$^{d}$, 
P. Ambrozewicz$^{e}$, 
D. Androic$^{f}$, 
T. Angelescu$^{g}$, 
R. Asaturyan$^{h}$, 
S. Avery$^{c}$, 
O.K. Baker$^{c,j}$, 
I. Bertovic$^{f}$,
H. Breuer$^{i}$,
R. Carlini$^{j}$, 
J. Cha$^{c}$, 
R. Chrien$^{k}$, 
M. Christy$^{c}$, 
L. Cole$^{c}$, 
S. Danagoulian$^{d}$, 
D. Dehnhard$^{l}$, 
M. Elaasar$^{m}$, 
A. Empl$^{b}$, 
R. Ent$^{j}$, 
H. Fenker$^{j}$, 
Y. Fujii$^{a}$, 
M. Furic$^{f}$, 
L. Gan$^{c}$, 
K. Garrow$^{j}$, 
A. Gasparian$^{c}$, 
P. Gueye$^{c}$, 
M. Harvey$^{c}$, 
O. Hashimoto$^{a}$, 
W. Hinton$^{c}$, 
B. Hu$^{c}$, 
E. Hungerford$^{b}$, 
C. Jackson$^{c}$, 
K. Johnston$^{o}$, 
H. Juengst$^{l}$, 
C. Keppel$^{c}$,
K. Lan$^{b}$, 
Y. Liang$^{c}$, 
V.P. Likhachev$^{p}$, 
J.H. Liu$^{l}$, 
D. Mack$^{j}$, 
A. Margaryan$^{h}$, 
P. Markowitz$^{q}$, 
J. Martoff$^{e}$, 
H. Mkrtchyan$^{h}$, 
S. N. Nakamura$^{a}$
T. Petkovic$^{f}$,
J. Reinhold$^{q}$, 
J. Roche$^{r}$,
Y. Sato$^{a,c}$, 
R. Sawafta$^{d}$, 
N. Simicevic$^{o}$, 
G. Smith$^{j}$, 
S. Stepanyan$^{h}$, 
V. Tadevosyan$^{h}$, 
T. Takahashi$^{a}$, 
K. Tanida$^{s}$,
L. Tang$^{c,j}$, 
M. Ukai$^{a}$, 
A. Uzzle$^{c}$, 
W. Vulcan$^{j}$, 
S. Wells$^{o}$, 
S. Wood$^{j}$, 
G. Xu$^{b}$, 
H. Yamaguchi$^{a}$, 
C. Yan$^{j}$\\
(HNSS Collaboration)}
\affiliation{
$^{a}$Tohoku University, Sendai, 980-8578, Japan; \\
$^{b}$University of Houston, Houston, TX 77204;\\ 
$^{c}$Hampton University, Hampton, VA 23668;\\
$^{d}$North Carolina A\&T State University, Greensboro, NC 27411;\\ 
$^{e}$Temple University, Philadelphia, PA 19122; \\
$^{f}$University of Zagreb, Zagreb, Croatia; \\
$^{g}$University of Bucharest, Bucharest, Romania;\\ 
$^{h}$Yerevan Physics Institute, Yerevan, Armenia; \\
$^{i}$University of Maryland, College Park, MD 20742; \\
$^{j}$Thomas Jefferson National Accelerator Facility, Newport News, VA 23606; \\
$^{k}$Brookhaven National Laboratory, Upton, NY 11973;\\
$^{l}$University of Minnesota, Minneapolis, MN 55455; \\
$^{m}$Southern University at New Orleans, New Orleans, LA 70126;\\
$^{n}$Rensselaer Polytechnic Institute, Troy, NY 12180;\\
$^{o}$Louisiana Tech University, Ruston, LA 71272; \\
$^{p}$University of Sao Paulo, Sao Paulo, Brazil; \\
$^{q}$Florida International University, Miami, FL 33199;\\
$^{r}$College of Williams and Mary, Williamsburg, VA 23187; \\
$^{s}$University of Tokyo, Tokyo 113-0033, Japan\\
}
\date{\today}
\begin{abstract}
High energy CW electron beams at new accelerator facilities
allow electromagnetic production and precision study of 
hypernuclear structure, and
we report here on the first 
experiment demonstrating the usefulness of the (e,e$^{\prime}$K$^{+}$)
reaction.  This experiment is also the first 
to take advantage of the enhanced virtual photon flux 
available when electrons are scattered at approximately 
zero degrees.  The observed resolution, $\sim$ 900 keV, of 
the $^{12}_{\Lambda}$B spectrum is
the best yet attained using magnetic spectrometers.  The positions of
the major excitations are in agreement with theoretical predictions
and the previous binding energy measurements.
\end{abstract}
\pacs{}
%
%
%
\maketitle

One of the goals of hypernuclear spectroscopy has been to determine an
effective $\Lambda$-Nucleon interaction\cite{millener1} which is then tested
against proposed $\Lambda$-Nucleon potentials folded into the nuclear
many-body problem.  A variety of spectroscopic data,
particularly data sensitive to the effective spin-dependent potential
parameters, is needed to help determine these
parameters\cite{millener2}, and this will improve the
theoretical treatment of the hadronic many-body system with
strangeness and the
knowledge of the free $\Lambda$-Nucleon interaction.
The (e,e$^{\prime}$K$^{+}$) reaction has a large spin-flip 
component, producing states
of non-natural parity, and 
substituting a $\Lambda$ for a proton in the target 
nucleus.  This produces hypernuclei charge symmetric to the previously
employed $(\mbox{K}^{-}, \pi^{-})$ and $(\pi^{+}, \mbox{K}^{+})$ 
reactions.

The cross section for electroproduction can be written in a very
intuitive form by separating out a factor, $\Gamma$, which multiplies
the off-shell (virtual) photoproduction cross section.  This factor
may be interpreted as the 
virtual photon flux produced by (e,e$^{\prime}$)
scattering\cite{barreau,hyde-wright,hungerford,xu}.  It has the form:

\begin{eqnarray}
\noindent \hspace*{0.5cm}\Gamma = \frac{\displaystyle 1}{\displaystyle
2 \pi} \, 
(\frac{\displaystyle \alpha}{\displaystyle -q_{u}^{2}})
\, \frac{\displaystyle\mbox{E}_{\gamma}}
{\displaystyle 1-\epsilon} \,  (\frac{\displaystyle \mbox{E}^{\prime}_{e}}
{\mbox{E}_{e}}).
\end{eqnarray}

\noindent In the above equation, $\epsilon$ is the polarization
factor which vanishes for scattering at zero degrees, and
$-q_{u}^{2}$ is the square of the
4-momentum of the virtual photon.
 
The virtual photoproduction cross section
can then be approximated by the
real photoproduction cross section at forward angles where the 
virtual photons are almost on the mass shell.
The virtual photon flux factor, $\Gamma$, also has the feature that it
is very forward peaked, so that for $\sim$ 2 GeV electron energies,
$\ge$ 50\% of
the total flux is captured between scattering angles of 
0 and 10mr\cite{hyde-wright,hungerford,xu}. 
Therefore if the electron spectrometer is placed at zero degrees, it
needs only a small acceptance to assure that almost all
the scattered electrons are transported to the focal plane, and 
that aberrations in the transport optics of the spectrometer
system will be small and easily corrected.
On the other hand, zero-degree-electroproduction uses essentially
transverse photons, and it results in very
high electron rates due to bremsstrahlung
in the target\cite{xu}.

The electron beam at Jlab is $\approx$ 60 $\mu$m in diameter with a
momentum spread of $\Delta$p/p $\approx 10^{-5}$.  At a beam
energy of $\sim$ 2 GeV, this introduces negligible error in the missing mass
resolution.  However, beam stability is a much more important issue, i.e. the
requirement that whatever the beam momentum, it must be stable and
reproducible for days (weeks) during the course of an experimental
run.  In the experiment reported here, beam stability was accomplished to the
level of a few parts in $10^{-4}$ by
implementing a feed-back lock to  the accelerator controls based on
momentum measurements of the beam in the arcs of the accelerator.
Changes in beam momentum were also recorded as each experimental run 
progressed, but corrections based on this information 
did not need to be applied.

The layout of the HNSS (E89-009) experiment is shown in Figure 1.  
An  electron beam of primary energy $\sim$ 1.8 GeV and 
$\sim 1\mu$A current, strikes a thin target (20 mg/cm$^{2}$)
placed just before a small zero-degree dipole magnet.  This magnet
splits the scattered particles;
the e$^{-}$ of about 300 MeV/c into a split-pole
spectrometer\cite{spencer}, ENGE, and
the K$^{+}$ of about 1.2 GeV/c into the Short Orbit Spectrometer, SOS,
which is fixed to the Hall C pivot at the Jefferson Laboratory.  
The splitting
magnet is necessary in order to capture the scattered electrons and
reaction kaons at very forward angles.

\begin{figure}[htb!]
\begin{center}
\epsfxsize = 8cm
\epsffile{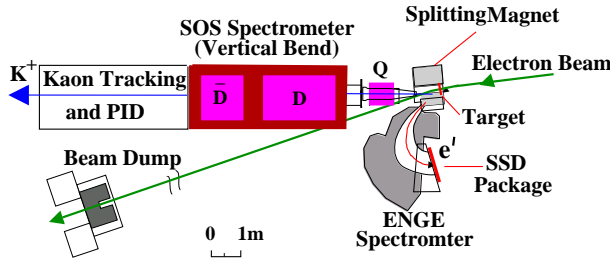}
\caption{The experimental plan view showing both the kaon
    spectrometer (SOS) and the electron spectrometer (ENGE)}
\label{fig:layout}
\end{center}
\end{figure}

The resolution for the experimental arrangement was expected
to be 600-1000 keV.  This was
dominated by the SOS spectrometer, which was not
designed for high resolution spectroscopy.
However, the SOS was a short-orbit magnet
(important as the decay length for a 1.2 GeV/c kaon was about 9m),
was already mounted and tested at the Hall C pivot, and had the 
sophisticated particle identification\cite{niculescu}, PID, package 
required to identify kaons within the expectedly large
background of pions and positrons.  Therefore, for the first experiment
demonstrating the feasibility of this experimental technique, the SOS
was chosen as the kaon spectrometer.

Two 
targets, 1) 22 mg/cm$^{2}$ $^{nat}$C, and 2) 19 mg/cm$^{2}$ $^{7}$Li, were
investigated, although a
calibration target, 8.8 mg/cm$^{2}$ CH$_{x}$ foil, was used to observe
$\Lambda$ and $\Sigma$ production from hydrogen.  Only the analysis of
the C and CH$_{x}$ data are reported here.  The reaction of
interest, (e,e$^{\prime}$K$^{+}$), resulted in the production of the
$^{12}_{\Lambda}$B hypernucleus when a C nucleus was targeted. 
Beam intensities were tuned to produce an acceptable
signal to accidental ratio, which for the C target was
approximately 0.6 $\mu$A, or an experimental luminosity of
$\sim 4 \times 10^{33}$ cm$^{-2}$-s$^{-1}$. A beam current of 0.4 $\mu$A
was used with the CH$_{x}$
target.  To satisfy conflicting beam requirements 
when simultaneously operating several experiments in the different
Halls, data had to be acquired at 
two different beam energies, 1721 and 1864 MeV.  However,
analysis has shown that within statistics, there are no differences
in the resulting $^{12}_{\Lambda}$B spectra.

Because of the high electron rates, the event trigger was the 
identification of a kaon in the SOS arm.
The coincidence of this kaon with an electron in the ENGE
spectrometer was completed in the off-line analysis. 
The number of positrons, pions, protons, and kaons per
second from the C target detected in the SOS was 
$ 10^{5}$, $1.4 \times 10^{3}$, 140, and 0.4
respectively. Therefore excellent particle identification was
required, not only in the analysis, but also in the hardware trigger.
The standard SOS detector package was used to identify
kaons, and its description and operation have been previously
discussed\cite{niculescu}.    The large flux of positrons in the SOS
came from  the acceptance of scattering angles at
$0^{\circ}$, where positrons from Dalitz pairs created in the target
were observed.  Positrons were easily identified and removed by
the shower counter PID, but were used here to confirm the experimental 
resolution.  

While the magnetic optics of the SOS were previously
well studied, it was necessary to recalibrate this
spectrometer because of the addition of the splitting magnet. For
example, the angular acceptance of the SOS, normally 15
msr, was reduced to 6 msr by this magnet.
The simulation of charged particle trajectories through this system of
magnets and detectors used the program,
RAYTRACE\cite{raytrace}.  The parameters of the RAYTRACE code
were determined by adjusting them so that the calculated
multi-dimensional particle distributions matched those observed when
the entrance angles and positions of reaction protons
and pions from the target were restricted by a
set of appropriately positioned holes in 
a tungsten plate (sieve slit) located between the splitting magnet and 
the SOS\cite{tang}. 
   
The electrons were detected near the focal plane of the Split-pole
spectrometer by a set of silicon strip detectors (SSD) specifically
designed for this experiment\cite{lan,miyoshi}.  Here only position
along the focal plane was needed to obtain the required resolution.
As it was also necessary to obtain a hit-position vs momentum
calibration for the ENGE spectrometer, RAYTRACE parameters
were adjusted to match the observed positions of the 
$p(e,e^{\prime} K^{+})\Lambda \, \mbox{or} \, \Sigma$
peaks produced from the CH$_{x}$ target. RAYTRACE also was used in a
Monte Carlo program to
determine the acceptance of the two spectrometers for the cross
section calculation.

The calibration spectrum obtained from the CH$_{x}$ target with the
accidental background, as taken with the 1864 MeV beam,
is shown in Figure 2.  The $\Lambda$ and $\Sigma$ peaks are obvious,
as well as an enhanced region just below the $\Lambda$ peak.  This
enhancement is
due to ($e,e^{\prime} K^{+}$) reactions on C in this target, and the 
threshold for
$^{12}_{\Lambda}$B production lies about 37.4 MeV below the $\Lambda$ peak.
The missing mass resolution in this spectrum, $\sim$ 3.5 MeV (FWHM), is 
dominated by the
kinematics due to the $\Lambda$ recoil, and cannot be corrected because
of the intrinsic angular resolution ($\sim$ 13mr FWHM) of the SOS 
spectrometer.  This kinematic effect falls
rapidly with target mass so that for the C target the contribution is
small.

\begin{figure}[htb!]
\begin{center}
\epsfxsize = 8cm
\epsffile{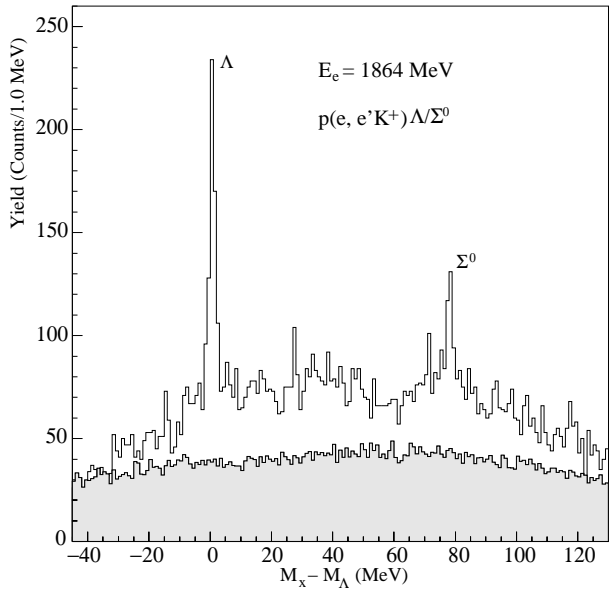}
\caption{The missing mass spectrum obtained 
from a CH$_{x}$ target at an incident electron energy of 1864 MeV.
The solid histogram is the accidental background as described in the text.}
\label{fig:cal_spec}
\end{center}
\end{figure}

Although the $\Lambda$ and
$\Sigma$ peaks in this spectrum were
used to calibrate the focal plane of the SSD detector
system, the known $\Lambda$ photoproduction cross section could not
be used for normalization because during irradiation the beam 
changed the hydrogen to carbon ratio of the CH$_{x}$ target.
The absolute value and shape of the accidental background was obtained by
averaging over 8 accidental time peaks.  

A measure of the experimental resolution can be obtained from the Dalitz
pair data, A($e,e^{\prime}[e^{+}e^{-}]$)A, where the two electrons are
observed in the ENGE and the positron in the SOS spectrometers, although
this reaction did not fill the entire acceptance of the 
spectrometer. In this spectrum, the projected full energy peak has an
experimental resolution of $\sim$900 keV, which predicts a system
resolution of $\sim$ 890 keV after removing the 
resolution (calculated) of one electron by quadratures.
This is consistent with the observed resolution
in the hypernuclear spectrum.

The $^{12}$C($e,e^{\prime} \mbox{K}^{+})^{12}_{\Lambda}$B spectrum 
with the accidental background is shown in figure 3. Clearly evident are
the $^{12}_{\Lambda}$B hypernuclear excitations where a $\Lambda$ is
inserted in the s
and p shells. The ground state doublet ($1^{-};2^{-}$) as well as the
($2^{+};3^{+}$) p-shell excitations are unresolved.
There is also strength in the bound-state region between these excitations,
although statistics are poor.

\begin{figure}[htb!]
\begin{center}
\epsfxsize = 8cm
\epsffile{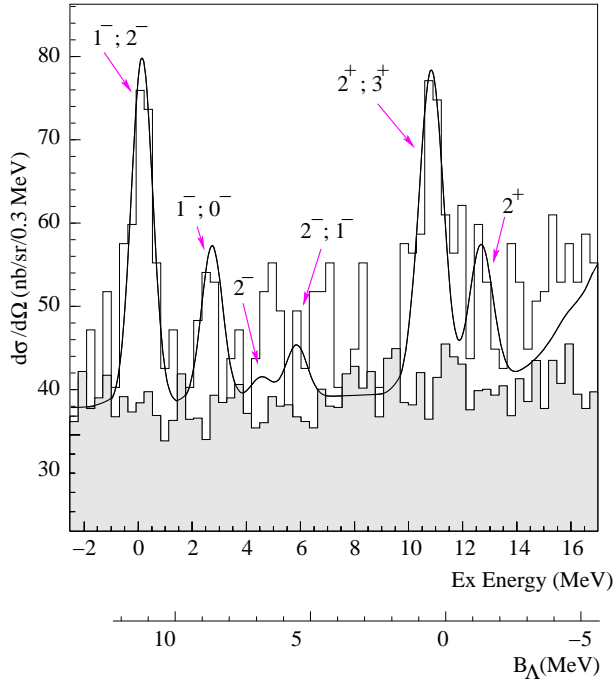}
\caption{\label{fig:hyp_spec}The summed $^{12}_{\Lambda}$B missing
mass spectrum for both energies.  The solid histogram is the measured
accidental background, and the curve is a theoretical calculation,
spread by 900 keV and overlayed on, not fit to, the data.}
\end{center}
\end{figure}

This spectrum is similar to
that predicted by Motoba, {\it et al} \cite{motoba1,motoba2} and by
Millener\cite{millener3}.  
Reference \cite{motoba2} also calculates
the excitation strengths for the photoproduction process of a kaon
at an angle of 3 deg by a 1.3 GeV photon.
The theoretical curve in the figure is
generated by superimposing Gaussian peaks, having a FWHM of 900
keV below and 5 MeV above 15 MeV excitation energy, 
on a polynomial fit to
the accidental background.  The strength of the peaks is taken
from reference \cite{motoba2} and their positions from
\cite{millener3}.  The 
Millener\cite{millener3} spectrum
used an effective $\Lambda$-Nucleus
interaction for the p-shell previously matched to other data, and 
thus should better represent the observed excitation energies.
The curve in the figure is directly overlayed on (not fitted to) the
data, without any shift in the energy scale.

The binding energy scale is determined from the position of the
$\Lambda$ and $\Sigma$ peaks in the calibration spectrum. The
$^{12}_{\Lambda}$B binding energy is found to be $11.4 \, \pm \,
0.5$ MeV compared to the emulsion measurement of 11.37 MeV.

This spectrum can also be compared to the
C($\pi^{+}$,K$^{+}$)$^{12}_{\Lambda}$C taken at KEK\cite{hotchi}.  The
resolution in this meson-induced spectrum is 1.45 MeV and shows 
the general shell structure as well as evidence for
core excited states, albeit with better statistics but poorer
resolution.  Of course this reaction excites natural parity
states, so although the overall shape of the spectrum is similar due to
the small value of the spin-dependent $\Lambda$N effective interaction, the
spin of the excited states differs.  We note in particular, the 
strength between 5 to 8 MeV excitation which is observed in both data.

The differential cross section can be calculated as if it were
photoproduction, by putting the virtual photons created in the
$(e,e^{\prime})$ reaction on the mass shell.  This in effect, averages
the elementary ($\gamma$,K) reaction at $\sim 1500$ MeV over the approximately
100 MeV spread of virtual gamma energies.   Since
various factors entering the cross section calculation changed as a
function of the incident energy, an independent analysis for
each energy was required.  The weighted average of these separate
cross section  
measurements is, $140 \, \pm \, 17(\mbox{stat}) \, \pm \, 18(\mbox{sys})$ 
nb/sr,  and is consistent with the value at both
energies separately.  It is also consistent with the theoretical 
prediction\cite{motoba2} of 152 nb/sr for the ground state.  

The cross section normalization can be compared for consistency with
two other experiments\cite{hinton,maeda}.  
Although these experiments did not resolve
hypernuclear bound states, they did determine the effective number of
protons participating in quasi-free production of
$\Lambda$s from a C nucleus.   To make this comparison, the
background subtracted quasi-free missing mass spectrum of our data
(not shown in the figure) was fitted 
by a polynomial function over a range of 23 MeV beginning at particle
threshold.  The ratio of the $\Sigma$ to the $\Lambda$ quasi-free 
production in this
spectrum was assumed to be the same as that extracted from \cite{hinton}.
To compare to the
previous experiments, the yield was corrected for acceptance, and 
differences in momentum transfer and energy\cite{bebek}. 
We obtain an effective quasi-free cross section of $4.55 \, \pm \, 0.19
\, \pm \, 0.46 \,\mu$b/sr compared to $4.2 \, \pm 0.4 \, \mu$b/sr,
\cite{maeda}. From this cross section,
we obtain 4.2 effective protons in quasi-free lambda
production from C, in agreement with previous experiments and
confirming our cross section normalization.
 
In summary, the first (e,e$^{\prime}$K$^{+}$) experiment has
successfully observed the electroproduction of
$^{12}_{\Lambda}$B. The missing mass resolution, $\sim900$ keV, is 
consistent with the expected value, and is a
factor of $\sim$ 1.6 better than existing
measurements.  The spectrum shape and differential cross sections are 
also consistent with theory, and the binding energy is consistent with
that previously measured by emulsion.  Although there is strength in the core
excited region, the statistical accuracy is not sufficient to
extract detailed structure information.  However the position and
strength of the states are similar to those
observed in the ($\pi^{+}$,K$^{+}$) reaction.
A normalization based on the measured quasi-free 
production is also consistent with previous
experiments\cite{hinton,maeda}, and confirms our cross section value.

\end{document}